\documentclass[aps,pra,reprint,superscriptaddress,amsmath,amssymb, floatfix]{revtex4-2}

\usepackage{graphicx}
\usepackage{dcolumn}
\usepackage{bm}
\usepackage{hyperref}

\begin{document}



\title{Broadband femtosecond lasers enable efficient two-photon excitation of the ultranarrow linewidth singlet 1s2s state in helium}


\author{Shashank Kumar}
\email{kumar414@purdue.edu}
\author{Justin D. Piel}%
\affiliation{Department of Physics and Astronomy, Purdue University, West Lafayette, Indiana 47906, USA}

\author{Chris H. Greene}
\author{Niranjan Shivaram}
\email{niranjan@purdue.edu}
\affiliation{Department of Physics and Astronomy, Purdue University, West Lafayette, Indiana 47906, USA}
\affiliation{Purdue Quantum Science and Engineering Institute, Purdue University, West Lafayette, Indiana 47907, USA}%

\date{\today}

\begin{abstract}
We propose a broadband, femtosecond two-photon excitation scheme for efficient population transfer to the ultra-narrow linewidth $1s2s\ ^1S_0$ metastable state in helium.  Using $120$ nm vacuum ultraviolet (VUV) femtosecond laser pulses, we theoretically demonstrate that a direct two-photon excitation process can achieve a population transfer efficiency of 25–30\%, even when photoionization losses are included.  The use of broadband pulses enables multiple excitation pathways to populate the excited state, in addition to compensating for significant AC Stark shifts occurring within the pulse duration. Furthermore, we introduce a two-color two-photon extreme ultraviolet – near infrared (XUV–IR) excitation scheme that will further reduce ionization losses and can achieve significantly higher transfer efficiencies of $\sim 70\%$. These results demonstrate that high excitation probability of ultra-narrow linewidth ($\sim 50$ Hz) excited states can be achieved with experimentally accessible femtosecond laser sources with a few THz bandwidth.
\end{abstract}

\maketitle


\section{\label{sec:introduction}Introduction}
Narrow-linewidth atomic states play a central role in quantum technologies, precision metrology, and advanced spectroscopy. Their long lifetimes allow coherent population control, a prerequisite for realizing long-lived qubits \cite{Shi2025, pucher2024, Harty2014, unnikrishnan2024, Wang2017, Wang2021} and high-precision atomic and nuclear clocks \cite{Aeppli2024, Bohman2023, Jefferts2014, Beeks2021, Zhang2024}. Among the most prominent examples are the metastable $2s\ ^2S_{1/2}$ state of atomic hydrogen (and hydrogen-like systems) and the $1s2s\ ^1S_0$ state of atomic helium, both of which exhibit lifetimes on the order of milliseconds \cite{shapiro1959, prior1972, kocher1972, hinds1978, van1971}. The longest-lived neutral atomic state is the first triplet excited state of helium, $1s2s\ ^3S_1$, which has a lifetime of $\sim 7800$ seconds \cite{hodgman2009}. Owing to their ultra-narrow linewidths, these states allow for extremely accurate determinations of transition frequencies. In particular, the $1s$–$2s$ transition in atomic hydrogen \cite{hildum1986, niering2000, parthey2011} and antihydrogen \cite{Ahmadi2018} has been measured with exceptional precision. These long-lived states primarily decay via two-photon emission \cite{lipeles1965, dalgarno1966, drake1969, bondy2020, prior1972}. A recent study suggests that the emitted photons could serve as a source of attosecond entangled biphotons \cite{wang2022}.

However, laser-assisted population transfer to a narrow-linewidth atomic state has long posed an experimental challenge. Narrow-bandwidth $\pi$-pulses are typically used to coherently populate these metastable states \cite{Ahmadi2018, grinin2020}, and similar approaches have been employed to excite atoms into Rydberg states \cite{mossberg1977, berman2007}. In practice, however, dynamic Stark shifts of the atomic levels and laser instabilities often limit the transfer efficiency. The requirement of narrow spectral bandwidths makes efficient transfer with a single pulse impractical, motivating the development of multipulse techniques. One widely established method is Stimulated Raman Adiabatic Passage (STIRAP), which employs a delayed pulse sequence to adiabatically transfer the population from the initial to the final state while avoiding population of the intermediate state \cite{gaubatz1990, Bergmann1998, Vitanov2017}. Another, theoretically proposed approach, is the Stark-chirped Rapid Adiabatic Passage (SCRAP), in which a multiphoton excitation pulse is combined with a delayed non-resonant pulse that induces a Stark shift to facilitate adiabatic transfer \cite{yatsenko1999, yatsenko2005, rickes2000}. Moreover, in many atomic and molecular systems, the ground-to-first-excited-state energy gap lies in the vacuum-ultraviolet (VUV) or extreme-ultraviolet (XUV) regime, where generating nanosecond pulses with sufficient power remains experimentally challenging, making efficient creation of excited states difficult in this regime.

In recent years, great advances have been made in generating VUV femtosecond pulses using techniques such as high-harmonic generation (HHG) \cite{Appi2020, konishi2020}, wave mixing processes \cite{misoguti2001, Forbes2024}, white-light supercontinuum formation \cite{dudley2006, Cheng2016}, and frequency-tunable resonant dispersive wave (RDW) emission \cite{im2010, joly2011, mak2013, belli2015, kottig2017, ermolov2015, Travers2019}. With broadband VUV pulses now readily available, two-photon excitation in large energy-gap systems, such as helium, has become experimentally feasible. However, the high peak intensities of femtosecond lasers (on the order of $\mathrm{TW/cm^2}$) inevitably drive strong multiphoton ionization, which severely limits population transfer efficiency. This presents a major hurdle for applications where high excitation probabilities of narrow excited states are required \cite{wang2022}.

In the present work, we investigate two-photon excitation in the helium atom using broadband femtosecond laser pulses, showing that such pulses can achieve high population transfer to the $1s2s$ $^1S_0$ state with ultra-narrow linewidth ($\sim 50~\mathrm{Hz}$), despite the large bandwidth ($\sim1~\mathrm{THz}$) and typically high multiphoton ionization rates of femtosecond pulses. In Sec.~\ref{sec:two_photon_120}, we derive an effective two-level Hamiltonian that depends on the dynamic Stark shifts, the two-photon Rabi frequency, and the ionization rate. Sec.~\ref{sec:numerical_and_plots} presents numerical solutions of this model for varying femtosecond pulse parameters (intensity and spectral bandwidth), from which we identify experimentally accessible regimes that maximize population transfer. In Sec.\ref{sec:two_color_two_photon}, we introduce a two-color, extreme-ultraviolet and near infrared (XUV–IR) two-photon excitation scheme, which results in high excitation probabilities of $\sim 70\%$. Finally, we conclude this article in Sec.~\ref{sec:Summary}.

\section{Theory of two-photon excitation}
\label{sec:two_photon_120}

\begin{figure}[h]
    \centering
    \includegraphics[width=0.9\linewidth]{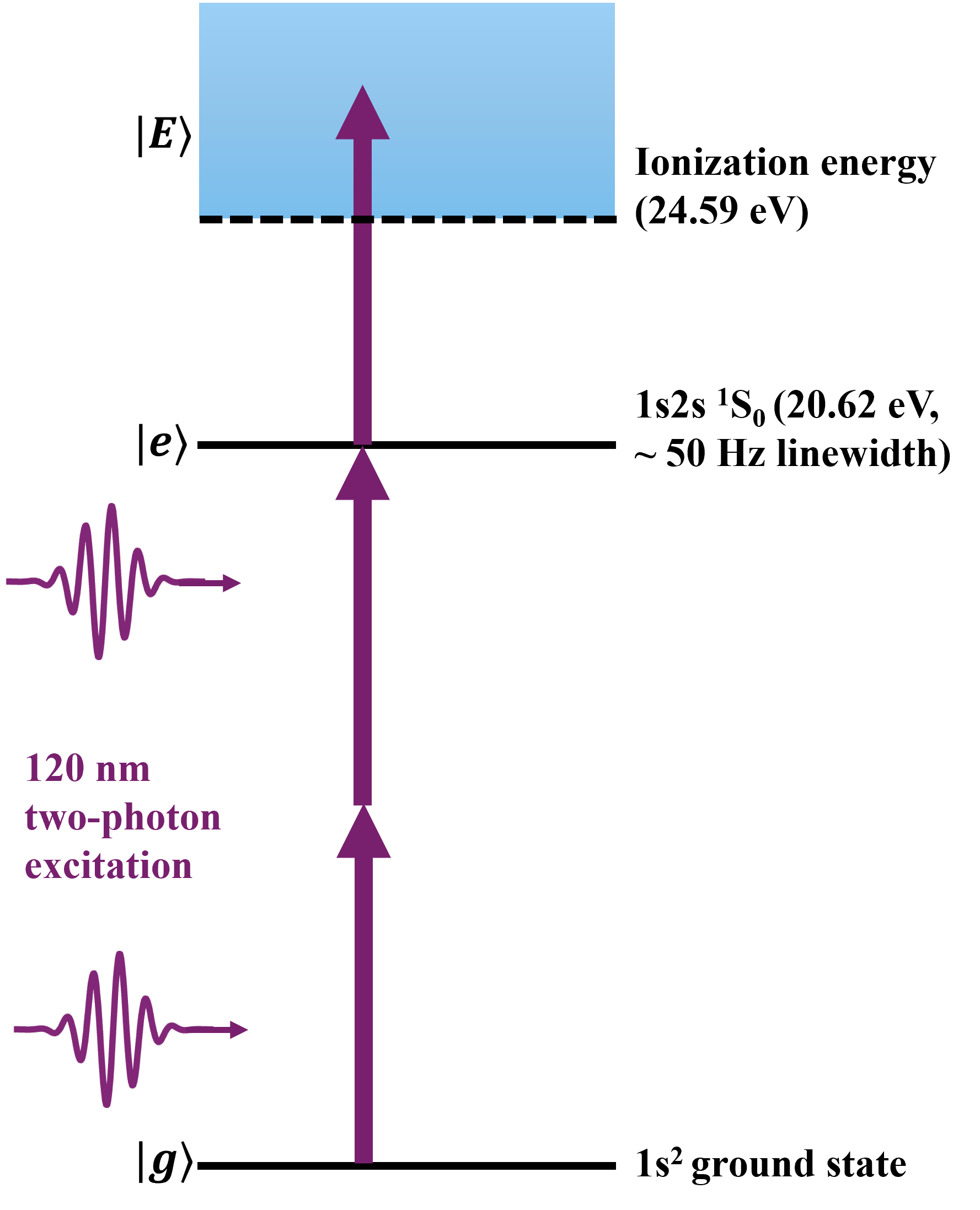}
    \caption{A two-photon excitation scheme with femtosecond 120 nm pulses including photoionization of the $1s2s$ $^1S_0$ state of the helium atom.}
    \label{fig:two_photon_ionization}
\end{figure}

\begin{figure*}[ht]
    \centering
    \includegraphics[width=\linewidth]{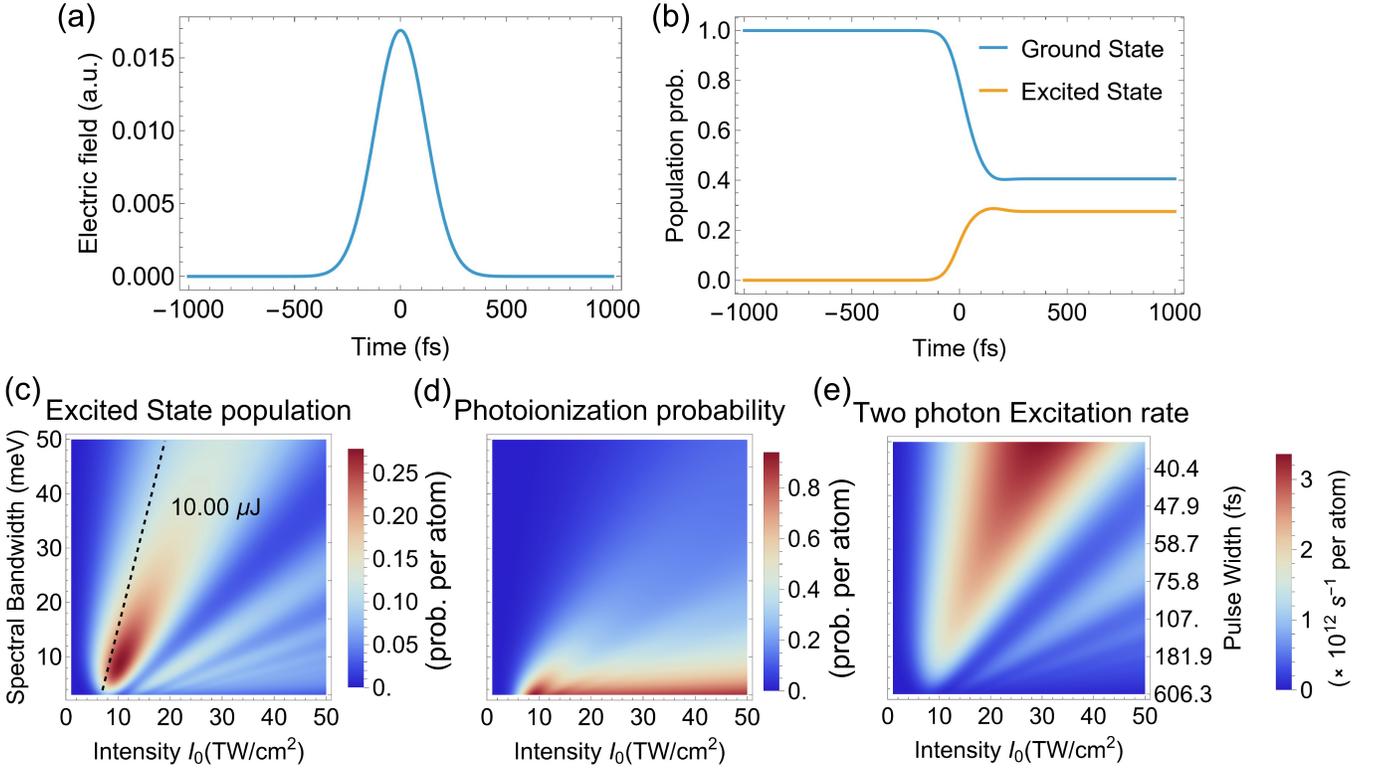}
    \caption{(a) The electric field profile of the pump (excitation) laser. (b) The time evolution of ground state (blue curve) and the excited state (orange curve) population corresponding to the pump laser electric field. (c) The excited state population $P_e(t=1000\text{ fs})$ is shown in the parameter space of intensity and pulse spectral bandwidth (full-width at half maximum). The dashed line is for constant pulse energy $=10 \mu J$. (e) The fraction of He atoms that are photoionized is $=1-|c_g|^2-|c_e|^2$. (f) The two-photon excitation rate. The Fourier transform limited (FTL) pulse width of the pump laser is labeled on the right side of the y-axis. The detuning in all the plots is $\Delta = 2\omega_P-\omega_{eg} = 14.3~meV$.}
    \label{fig:two_photon_120nm}
\end{figure*}

\begin{figure*}[ht]
    \centering
    \includegraphics[width=0.9\linewidth]{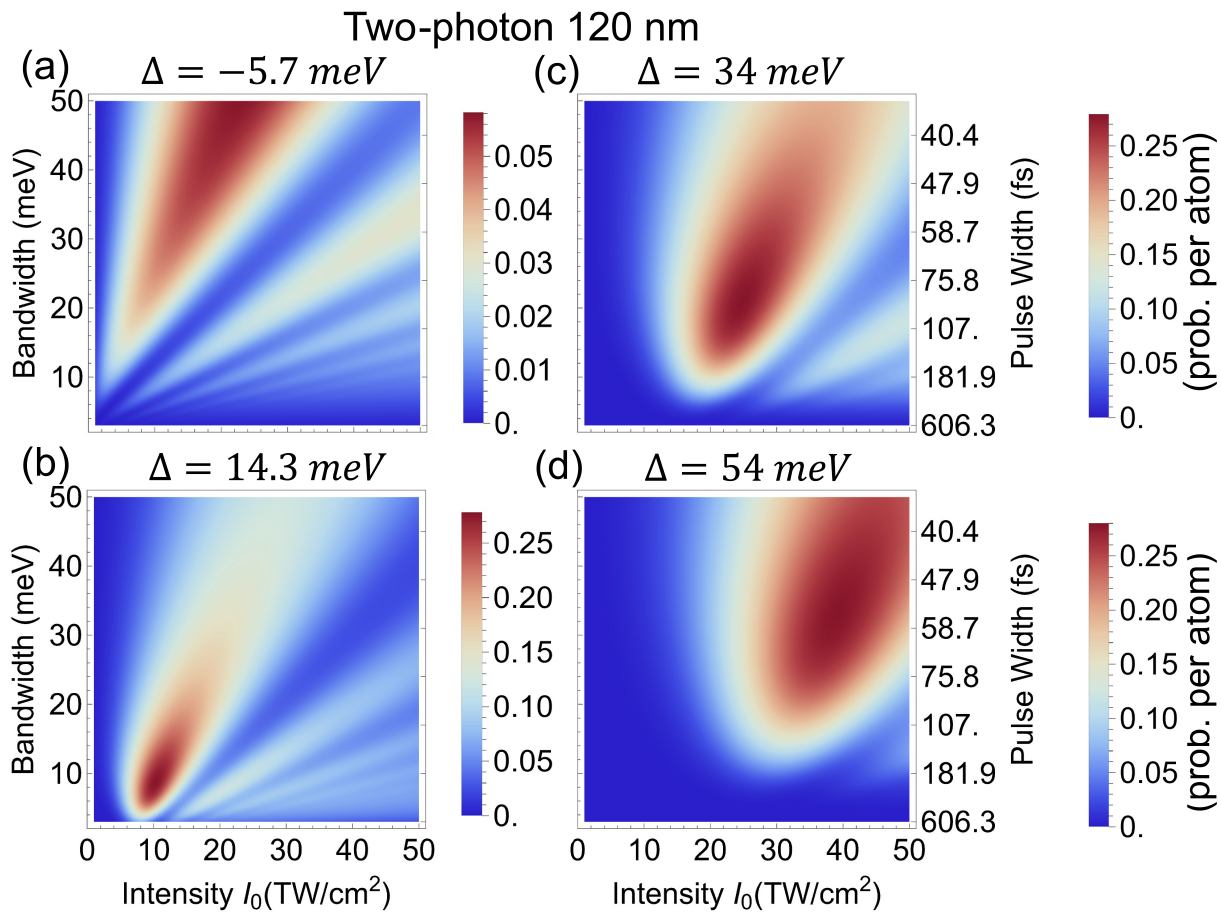}
    \caption{The excitation probability as a function of pump intensity and pulse width for various two-photon detuning $\Delta=2\omega_P - \omega_{eg}$. (a) $\Delta = -5.7$ meV (b) $\Delta = 14.3$ meV (c) $\Delta = 34$ meV (d) $\Delta = 54$ meV.}
    \label{fig:two_photon_detuning}
\end{figure*}

In this section, we present the theoretical framework for two-photon excitation of a helium atom by a broadband pulse. The system consists of the ground state $1s^2$ and the first excited state $1s2s\ ^1S_0$ of the helium atom, separated by $20.62\ \mathrm{eV}$. Because both states have the same parity ($l=0$), a single-photon dipole transition is forbidden, and excitation requires the absorption of two photons (see Fig.~\ref{fig:two_photon_ionization}). Since our primary focus is on the final population of the $1s2s$ state, the framework also incorporates photoionization pathways that compete with the two-photon excitation process.
The total Hamiltonian is written as $H = H_0 + V(t)$, where $H_0$ is the field-free atomic Hamiltonian. Its eigen energies and eigen functions are obtained using finite-volume variational $R$-matrix methods \cite{Aymar1996}, which are well established to accurately describe helium excitation and ionization continua. The laser–atom interaction is treated within the dipole approximation, $V(t) = -\bm{\mu}\cdot\bm{E}_P(t)$, where $\bm{\mu}$ is the dipole operator and $\bm{E}_P(t)$ is the pump electric field. For a femtosecond laser pulse, the field takes the form
\begin{equation}
    \bm{E}_P(t)=\bm{\varepsilon}_P(t)\cos{(\omega_Pt)}
\end{equation}
with $\bm{\varepsilon}_P(t)$ denoting the Gaussian envelope and $\omega_P$ the central angular frequency. The temporal phase is assumed to be zero. In Fig.~\ref{fig:two_photon_ionization}, we consider a pulse with central wavelength $120.27\ \mathrm{nm}$, corresponding to $\hbar \omega_P = 10.31\ \mathrm{eV}$, such that two-photon absorption matches the $1s^2 \to 1s2s$ excitation. Later, we will also analyze a two-color excitation scheme, where the two absorbed photons have different frequencies. Unless stated otherwise, the following Gaussian intensity profile is used throughout this article.
\begin{equation}
\label{int_prof}
    I(t)=I_0\exp{\left(-4\ln{2}\frac{t^2}{\tau^2}\right)}
\end{equation}
$\tau$ is the full width at half maximum and $I_0$ is the peak intensity of the excitation pulse. The envelope function of the electric field can be determined by using the relation $I=\frac{1}{2}\epsilon_0c\varepsilon_P^2$.

The total time-dependent wavefunction for such a system can be written as
\begin{align}
\label{wavefunction}
    |\psi\rangle=&c_{g} e^{-i \omega_{g} t}|g\rangle+c_{e} e^{-i \omega_{c} t}|e\rangle \notag\\
    &+ \sum_{m} c_{m} e^{-i \omega_{m} t}|m\rangle+\int dE\ c_{E}(t) e^{-i \omega_{E} t}|E\rangle
\end{align}
Here, $|g\rangle$ denotes the ground state, $|e\rangle$ the excited state $1s2s\ ^1S_0$, and $|E\rangle$ the continuum states. $|m\rangle$ represents intermediate $p$ states that mediate the two-photon excitation, but are omitted for the sake of clarity in Fig.~\ref{fig:two_photon_ionization}. Substituting Eq.~\ref{wavefunction} into the time-dependent Schrödinger equation yields the coupled rate equations for the complex amplitudes $c_{i}(t)$.

\begin{subequations}
\label{eq_of_motion}
    \begin{align}
        i \hbar \dot{c}_{g}(t)=&\sum_{m} c_{m}(t) e^{-i \omega_{mg} t} V_{g m}(t)\\
        i \hbar \dot{c}_{e}(t)=&\sum_{m} c_{m}(t) e^{-i \omega_{m e} t} V_{e m}(t)\notag\\
        &+\int d E c_{E}(t) e^{-i \omega_{e E} t} V_{e E}(t)\\
        i \hbar \dot{c}_{m}(t)=&\sum_{j=g,e} c_{j}(t) e^{i \omega_{mj} t} V_{mj}(t)\\
        i \hbar \dot{c}_{E}(t)=&c_{e}(t) e^{i \omega_{E e} t} V_{E e}(t)
    \end{align}
\end{subequations}
where, $V_{ij}(t) = -\langle i|\bm{\mu}\cdot\bm{E}_P(t)|j\rangle$ is the dipole interaction matrix element and $\omega_{ij}= \omega_i - \omega_j$. Since the transition to intermediate states $|m\rangle$ is off resonant, the highly oscillating time-dependent amplitudes can be evaluated using adiabatic elimination \cite{kaufman2020}. Eq.~\ref{eq_of_motion} (c) can then be integrated to obtain
\begin{align}\label{intermediate_state}
    c_{m}(t)\approx\sum_{j=g,e} \frac{c_{j} \mu_{m j}^{P}}{2 \hbar}\left[\frac{\varepsilon_{P} e^{-i\left(\omega_{P}-\omega_{m j}\right) t}}{\omega_{m j}-\omega_{P}}+\frac{\varepsilon_{P}^{*} e^{i\left(\omega_{P}+\omega_{m j}\right) t}}{\omega_{m j}+\omega_{P}}\right]
\end{align}
where $\mu_{mj}^P = \langle m|\bm{\mu}\cdot\hat{e}_P|j\rangle$ is the dipole matrix element projected along the polarization direction of the pump pulse. Similarly, Eq.~\ref{eq_of_motion} (d) governs the time evolution of the continuum amplitudes.
\begin{align}\label{continuum_amp}
    c_{E}(t)=\frac{-1}{2 i \hbar} &\int_{-\infty}^{t} d t^{\prime} c_{e}\left(t^{\prime}\right) \mu_{E e}^{P}\notag\\
    &\times \left[\varepsilon_{P} e^{-i\left(\omega_{P}-\omega_{E e}\right) t}+\varepsilon_{P}^{*} e^{i\left(\omega_{P}+\omega_{E e}\right) t^{\prime}}\right]
\end{align}
Now, substituting these amplitudes into Eqs.~\ref{eq_of_motion} (a) and (b), we obtain the coupled differential equations for the amplitudes $c_g(t)$ and $c_e(t)$. Within the rotating-wave approximation (RWA) and the two-photon RWA, where rapidly oscillating terms beyond the two-photon detuning $\Delta = 2\omega - \omega_{eg}$ are neglected. This simplifies to
\begin{equation}
\label{coupled_eqn}
    i
    \begin{pmatrix}
        \dot{c}_g\\
        \dot{c}_e
    \end{pmatrix}
    =
    \begin{pmatrix}
        S_g(t) & \frac{\Omega_{ge}^{P*}}{2}e^{i\Delta t}\\
        \frac{\Omega_{ge}^{P}}{2}e^{-i\Delta t} & S_e(t) + \frac{\delta E_e}{\hbar} -\frac{i}{2}\Gamma_e^P 
    \end{pmatrix}
    \begin{pmatrix}
        c_g\\
        c_e
    \end{pmatrix}
\end{equation}
Here, $\hbar S_{g/e}$ is the dynamical stark shift of the ground / excited state and is given by,
\begin{equation}
\label{stark_shift}
    S_{j}(t) = -\sum_{m} \frac{\left|\mu_{m j}^{P}\right|^{2}}{2\hbar^{2}}\left|\varepsilon_{P}\right|^{2} \frac{\omega_{m j}}{\omega_{m j}^{2}-\omega_{P}^{2}}
\end{equation}
The effective two-photon Rabi frequency coupling the ground state and the excited state can be expressed as
\begin{equation}
\label{rabi_freq}
    \Omega_{ge}^{P} = -\sum_{m}\frac{\mu_{e m}^{P} \mu_{m g}^{P}}{2 \hbar^{2}} \frac{\varepsilon_{P}^{2}(t)}{\omega_{m g}-\omega_{P}}
\end{equation}
The term $\delta E_e$ represents the dynamical energy shift of the $1s2s$ state arising from its coupling to the continuum, while $\Gamma_e^P$ denotes the photoionization rate of the excited state $1s2s$ \cite{KNIGHT1990, toth2023}.
\begin{align}
    \label{stark_shift_continumm}
    \delta E_e &= - \mathcal{P}\int d E \frac{1}{4\hbar}\frac{\left|\mu_{e E}^{P}|^2 |\varepsilon_{P}\right|^{2}}{\omega_E - \omega_P - \omega_e}\\
    \label{photoionization}
    \Gamma_{e}^P(t)&=\frac{\pi}{2 \hbar} \left|\varepsilon_{P}\right|^{2}\left|\left\langle e\left| \mu \cdot \hat{e}_{P}\right| E=\hbar \omega_{P}+\hbar \omega_{e}\right\rangle\right|^{2}
\end{align}
where $\mathcal{P}$ indicates the principal value.
\section{Time evolution of population}
\label{sec:numerical_and_plots}

After applying adiabatic elimination, Eq.~\ref{coupled_eqn} is solved numerically using the explicit Runge–Kutta method. Fig.~\ref{fig:two_photon_120nm} (a) shows the electric field envelope obtained from Eq.~\ref{int_prof}. Using this field, we compute the time evolution of the ground and excited state amplitudes of the helium atom. The corresponding state populations, $P_g = |c_g|^2$ and $P_e = |c_e|^2$, are presented in Fig.~\ref{fig:two_photon_120nm} (b).

The excited-state population is evaluated over a range of laser intensities and pulse bandwidths, with the results shown in Fig.~\ref{fig:two_photon_120nm} (c) as a map in the intensity–bandwidth parameter space. These calculations indicate that a population transfer efficiency of approximately 25\% can be achieved with a laser pulse intensity of $\sim 10\ \text{TW/cm}^2$ and a pulse bandwidth of $9\ \mathrm{meV}$ (the corresponding pulse full-width at half maximum is $\sim 200\ \text{fs}$). These parameters are experimentally accessible with table-top femtosecond laser systems using, for example, the RDW technique \cite{im2010, joly2011, mak2013, belli2015, kottig2017, ermolov2015, Travers2019}. The dashed line in Fig.~\ref{fig:two_photon_120nm} (c) marks the contour corresponding to an experimentally feasible pulse energy of $10\ \mu\text{J}$ focused to a $50\ \mu\text{m}$ spot diameter. However, at such high intensities, photoionization starts becoming a limiting factor, reducing the excited-state population particularly for long pulses, where the long time light–atom interaction further enhances ionization losses, as shown in Fig.~\ref{fig:two_photon_120nm} (d). We also evaluate the two-photon excitation rate, defined as the excitation probability divided by the full-width at half-maximum (FWHM) pulse width. As shown in Fig.~\ref{fig:two_photon_120nm} (e), for the intensity and bandwidth parameters specified above, the excitation rate is $\sim 2 \times 10^{12}\ \text{s}^{-1}\text{atom}^{-1}$.

Broadband laser pulses have been previously employed for efficient multiphoton excitation processes \cite{Nobis2023, meshulach1999}. Their large spectral bandwidth provides multiple excitation pathways, thereby enhancing the overall multiphoton transition probability. The two-photon excitation probability can be calculated using second-order perturbation theory and is proportional to the self-convolution of the spectrum of the broadband fields.
\begin{equation}
\label{convolution}
    P_{eg} \propto \left|\int d\omega E(\omega_{eg}-\omega ) E(\omega_{eg}+\omega )\right|^2
\end{equation}
Equation~\ref{convolution} indicates that the two-photon excitation probability arises from the coherent sum of all possible frequency pathways within the pulse bandwidth. As a result, the excitation probability remains remarkably high for appropriately chosen values of detuning, intensity, and bandwidth, as illustrated in Fig.~\ref{fig:two_photon_detuning}.


\section{XUV-IR two-color Scheme}
\label{sec:two_color_two_photon}

\begin{figure}[t]
    \centering
    \includegraphics[width=0.9\linewidth]{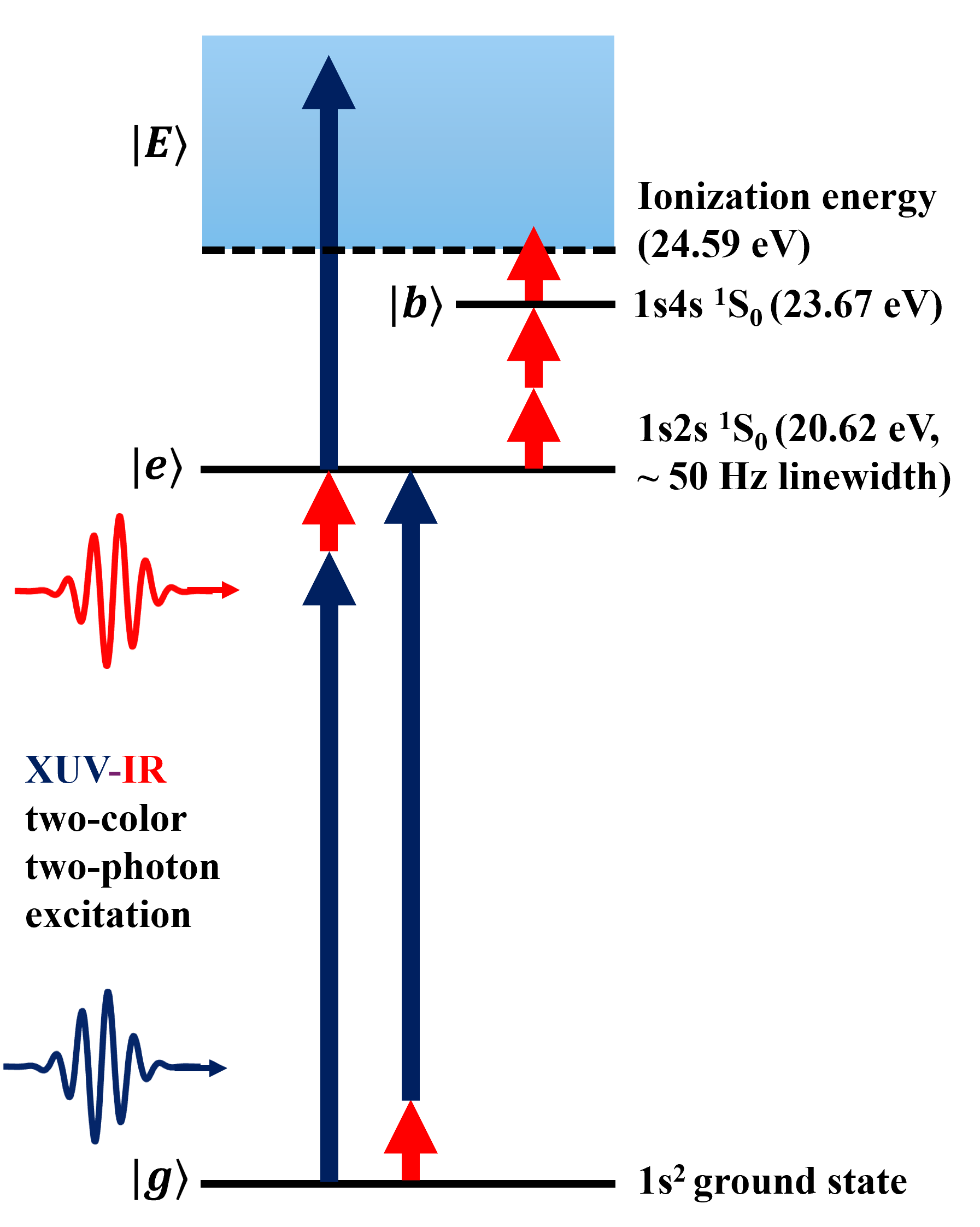}
    \caption{Two-color two-photon excitation scheme in a helium atom. There are two possible combinations of XUV and IR photons shown, which lead to two-photon excitation. IR photon energy is selected such that it is lower than the energy needed to one-photon photoionize the $1s2s$ state. However, three-photon ionization is possible with a two-photon resonance between $1s2s$ and $1s4s$ states as shown.}
    \label{fig:two_color_scheme}
\end{figure}

XUV femtosecond laser pulses have been generated using free electron lasers (FELs) for several decades \cite{MURPHY1985, Radcliffe2012}. Recent advances in FEL technology have substantially increased the photon flux of these pulses \cite{Finetti2017, Bidhendi2022}, making them highly suitable for multiphoton excitation experiments.

In the previous section, we examined the one-color ($120\ \text{nm}$) two-photon excitation scheme, which, despite significant photoionization, demonstrated efficient population transfer in helium. Here, we propose a two-color approach to further reduce the photoionization rate. In this scheme, a broadband XUV pulse is employed in combination with a near-IR pulse for two-color two-photon excitation, as illustrated in Fig.~\ref{fig:two_color_scheme}. The IR pulse is chosen so that it lacks sufficient energy to directly ionize the $1s2s$ state. However, multiphoton ionization of the excited state can still occur via the $1s4s$ intermediate state, as also indicated in Fig.~\ref{fig:two_color_scheme}. Since the photoionization cross section generally decreases with increasing photon energy \cite{Rafiq2007, reitsma2019}, the photoionization rate of the $1s2s$ state due to the XUV pulse, calculated using Eq.~\ref{photoionization}, is an order of magnitude smaller than that for the 120 nm pulse.

\begin{figure*}[ht]
    \centering
    \includegraphics[width=0.9\linewidth]{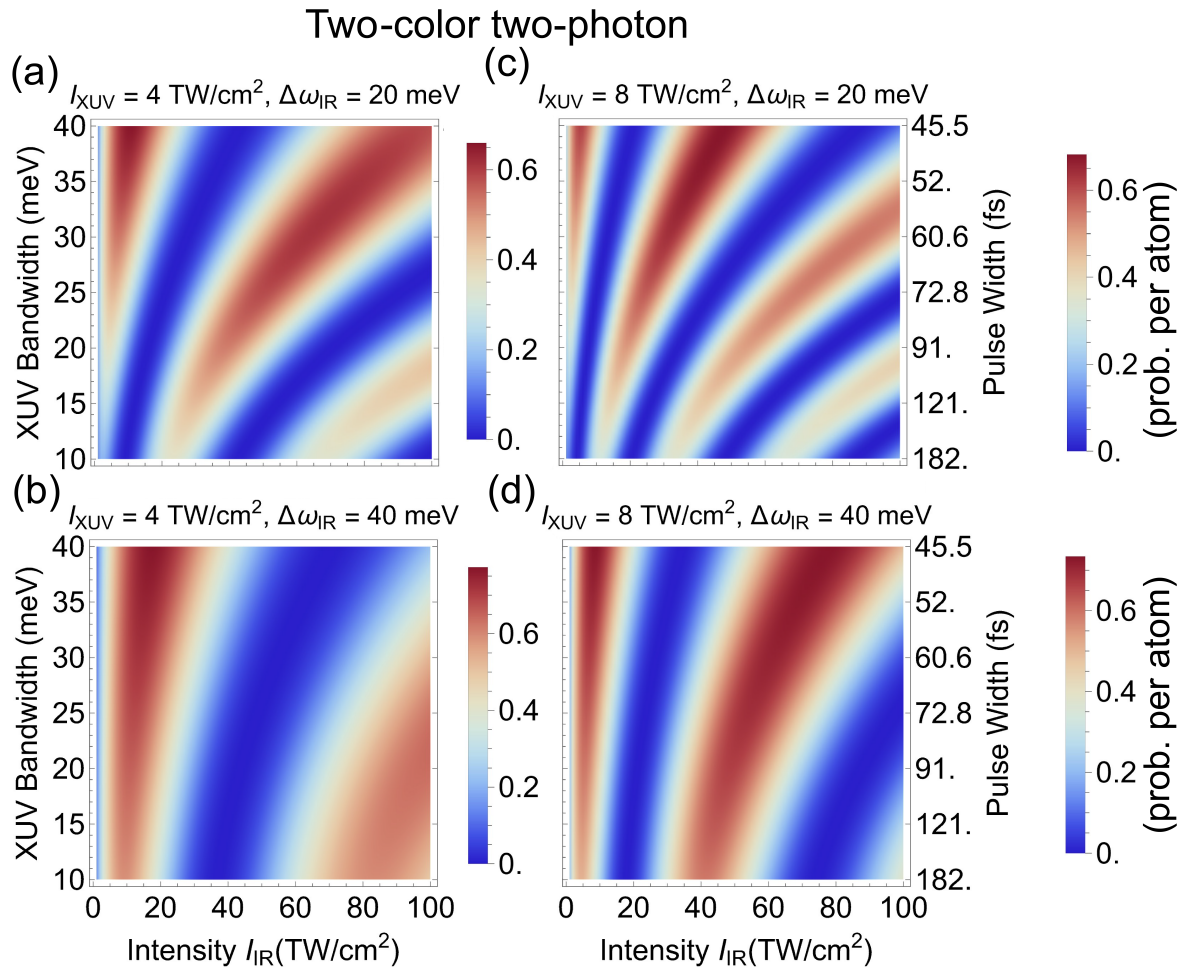}
    \caption{The excitation probability as a function of IR pulse ($\omega_2$) intensity and width for XUV pulse (a) XUV Intensity = $4\ \mathrm{TW/cm^2}$ and IR bandwidth = $20$ meV (pulse width $\sim 91$ fs) (b) XUV Intensity = $4\ \mathrm{TW/cm^2}$ and IR bandwidth = $40$ meV (pulse width $\sim 45$ fs) (c) XUV Intensity = $8\ \mathrm{TW/cm^2}$ and IR bandwidth = $20$ meV (pulse width $\sim 91$ fs) (d) XUV Intensity = $8\ \mathrm{TW/cm^2}$ and IR bandwidth = $40$ meV (pulse width $\sim 45$ fs).}
    \label{fig:two_color}
\end{figure*}

For this scheme, we can modify the wavefunction, Eq.~\ref{wavefunction} as,
\begin{align}
\label{wavefunction_3level}
    |\psi\rangle=&c_{g} e^{-i \omega_{g} t}|g\rangle+c_{e} e^{-i \omega_{c} t}|e\rangle + c_be^{-i\omega_b t}|b\rangle \notag\\
    &+ \sum_{m} c_{m} e^{-i \omega_{m} t}|m\rangle+\int dE\ c_{E}(t) e^{-i \omega_{E} t}|E\rangle
\end{align}
Here, $|b\rangle$ denotes the state $1s4s\ ^1S_0$, which is two-photon resonant with the state $1s2s\ ^1S_0$ for the IR pulse. Following the derivation in Sec.~\ref{sec:two_photon_120}, the evolution of the coefficients can be written within the RWA and two-photon RWA as
\begin{widetext}
\begin{equation}
\label{coupled_eqn_two_color}
    i
    \begin{pmatrix}
        \dot{c}_g\\
        \dot{c}_e\\
        \dot{c}_b
    \end{pmatrix}
    =
    \begin{pmatrix}
        S_g^1(t)+S_g^2(t) & (\frac{\Omega_{ge}^{21}}{2} + \frac{\Omega_{ge}^{12}}{2})^*e^{i\Delta_{eg} t} & 0\\
        (\frac{\Omega_{ge}^{21}}{2} + \frac{\Omega_{ge}^{12}}{2})e^{-i\Delta_{eg} t} & S_e^1(t) + S_e^2(t) + \frac{\delta E_e^1}{\hbar} -\frac{i}{2}\Gamma_e^1 & \frac{\Omega_{be}^*}{2}e^{i\Delta_{be}t}\\
        0 & \frac{\Omega_{be}}{2}e^{-i\Delta_{be}t} & S_b^1(t) + S_b^2(t) + \frac{\delta E_b^2}{\hbar} -\frac{i}{2}\Gamma_b^2
    \end{pmatrix}
    \begin{pmatrix}
        c_g\\
        c_e\\
        c_b
    \end{pmatrix}
\end{equation}
\end{widetext}
where $S_{g/e/b}^{1(2)}$ denote the Stark shifts of the ground, $1s2s\ ^1S_0$, and $1s4s\ ^1S_0$ states induced by pulses $\omega_1$ and $\omega_2$, corresponding to the XUV and IR photon energies, respectively. The two-photon Rabi frequencies $\Omega_{ge}^{12}$ and $\Omega_{ge}^{21}$ correspond to the two possible excitation pathways depicted in Fig.~\ref{fig:two_color_scheme}, while $\Omega_{be}$ represents the two-photon Rabi frequency for the $1s2s \to 1s4s$ transition. The two-photon detunings are defined as $\Delta_{eg} = (\omega_1 + \omega_2) - \omega_{eg}$ and $\Delta_{be} = 2\omega_2 - \omega_{be}$. Since IR ($\omega_2$) does not have sufficient energy to photoionize the $1s2s$ state, the energy shift $\delta E_e^1$ and the photoionization rate $\Gamma_e^1$ are associated with the XUV ($\omega_1$) pulse only. In contrast, the IR pulse can photoionize the $1s4s$ state, so $\delta E_b^2$ and $\Gamma_b^2$ correspond to the IR-induced effects. Single-photon ionization of the $1s4s$ state by the XUV pulse is neglected because it is orders of magnitude smaller than that induced by the IR pulse. For detailed expressions of these quantities, see Appendix~\ref{appendix_B}

For the numerical calculations, we chose $\hbar\omega_1 = 19.06$ eV ($\sim 65 $ nm, XUV) and $\hbar\omega_2 = 1.54$ eV ($\sim 800$ nm, IR). Fig.~\ref{fig:two_color} shows the excited state population as a function of IR pulse intensity and pulse width for various values of XUV intensities and pulse widths. This two-color scheme demonstrates that, with appropriately chosen pulse parameters, population transfer efficiencies of up to $70\%$ can be achieved. The populations shown in Fig.~\ref{fig:two_color} correspond to pulse parameters that are currently experimentally accessible at the FLASH FEL \cite{Bidhendi2022}.

\section{Summary}
\label{sec:Summary}

In conclusion, we have presented a detailed theoretical study of two-photon excitation in helium using broadband femtosecond laser pulses to enhance population transfer efficiency to an ultra-narrow linewidth state. We showed that the presence of multiple excitation pathways and the broad spectral bandwidth can compensate for both the dynamical Stark shift and the large photoionization rates, thereby significantly increasing the transfer efficiency from the ground state to the $1s2s$ state. For the 120 nm two-photon excitation scheme, we mapped the excited-state population in the intensity–bandwidth parameter space and found that populations $>25\%$ are achievable with experimentally accessible laser parameters. Finally, we proposed a two-color two-photon excitation scheme employing an XUV and an IR pulse. Our calculations show that, under experimentally realizable parameters, this scheme can achieve excited-state populations of $\sim70\%$. This approach opens new avenues for efficiently populating ultra-narrow linewidth atomic states with broadband lasers, paving the path for simplified and more versatile control schemes beyond traditional adiabatic passage techniques such as STIRAP~\cite{gaubatz1990, Bergmann1998, Vitanov2017} and SCRAP\cite{yatsenko1999, yatsenko2005, rickes2000}. In the case of helium atoms, as demonstrated in this study, and in helium-like systems, such a high excitation probability of the 1s2s $^1S_0$ state could lead to the generation of attosecond bandwidth entangled photons with high rates \cite{wang2022}. 


\begin{acknowledgments}
We gratefully acknowledge support from the W.M. Keck Foundation. We thank Prof. Francis Robicheaux, Purdue University, for insightful discussions.
\end{acknowledgments}

\appendix

\section{Derivation of effective Hamiltonian for two-photon 120 nm excitation scheme}
\label{appendix_A}

The time-dependent Schrodinger equation with laser-atom interaction can be written as,
\begin{equation}\label{time-dep-eq}
    \frac{\partial |\psi\rangle}{\partial t} = (H_0-\bm{\mu}\cdot \bm{E_p}(t))|\psi\rangle
\end{equation}
where, $H_0$ is the field-free Hamiltonian of the helium atom and $-\bm{\mu}\cdot \bm{E_p}(t)$ is the time-dependent Hamiltonian within the dipole approximation. Putting in Eq.~\ref{wavefunction}, the above time-dependent Schrodinger equation becomes Eq.~\ref{eq_of_motion}. The energy difference between state $m$ and $j$ is $\omega_{mj} = \omega_m - \omega_j$ and $V_{mj}$ are the elements of the interaction matrix and are defined as
\begin{align}
    V_{mj} &= \langle m \lvert V(t)\lvert j\rangle =-\sum_{j} \frac{\mu_{m j}^{P}}{2}\left(\varepsilon_{P} e^{-i\omega_{P}t}+\varepsilon_{P}^{*} e^{i\omega_{P}t}\right)
\end{align}
where $\mu_{mj}^{P} = \langle m \lvert \bm{\mu} \cdot \bm{e}_{P}\lvert j\rangle$ are the transition dipole matrix elements. The off-resonant states $c_m(t)$ are calculated by integrating Eq~\ref{coupled_eqn} (c) both sides,
\begin{align}
    &c_j(t) \notag\\
    &= -\sum_{j} \frac{c_{j} \mu_{m j}^{P}}{2}\int dt\left(\varepsilon_{P} e^{-i\left(\omega_{P}-\omega_{m j}\right) t}+\varepsilon_{P}^{*} e^{i\left(\omega_{P}+\omega_{m j}\right) t}\right)
\end{align}
Using adiabatic elimination, this coefficient can be approximated in the analytical form shown in Eq.~\ref{intermediate_state}. The adiabatic elimination approximation has previously been shown to be valid for broadband femtosecond pulses \cite{kaufman2020}. Substituting Eq.~\ref{intermediate_state} in in the Eq~\ref{eq_of_motion} (a),
\begin{align}
    &i\hbar\dot{c}_g(t) \notag \\
    &= \sum_m \sum_{j} \frac{c_{j} \mu_{m j}^{P}}{2 \hbar}\left[\frac{\varepsilon_{P} e^{-i\left(\omega_{P}-\omega_{m j}\right) t}}{\omega_{m j}-\omega_{P}}+\frac{\varepsilon_{P}^{*} e^{i\left(\omega_{P}+\omega_{m j}\right) t}}{\omega_{m j}+\omega_{P}}\right] \notag\\
    &\times e^{-i\omega_{mg}t} \left[-\frac{\mu_{g m}^{P}}{2}\left(\varepsilon_{P} e^{-i \omega_{P} t}+\varepsilon_{P}^{*} e^{i \omega_{P} t}\right)\right]
\end{align}
Using RWA and Two-photon RWA (TPRWA), the terms that oscillate faster than the two-photon detuning $\Delta = 2\omega_P - \omega_{eg}$ to get,
\begin{align}
    =-\sum_{m} c_{g}(t)\frac{\mu_{m g}^{P} \mu_{g m}^{P}}{4 \hbar}\left[\frac{\left|\varepsilon_{P}\right|^{2}}{\left(\omega_{m g}-\omega_{P}\right)}+\frac{\left|\varepsilon_{P}\right|^{2}}{\left(\omega_{m g}+\omega_{P}\right)}\right]\notag\\
    -\sum_{m} c_{e}(t) \frac{\mu_{mg}^{P} \mu_{gm}^{P}}{4\hbar} \frac{\varepsilon_{P}^{*2}(t)}{\omega_{m e}+\omega_{p}} e^{-i\left(\omega_{e g}-2 \omega_P\right) t}
\end{align}
\begin{align}\label{ground_amp}
    i \dot{c}_{g} \approx -\sum_{m} \sum_{P} \frac{\left|\mu_{m g}^{P}\right|^{2}}{2 \hbar^{2}}\left|\varepsilon_{P}\right|^{2} \frac{\omega_{m g}}{\omega_{m g}^{2}-\omega_{P}^{2}}c_{g}(t)\notag\\
    -\sum_{m}\frac{\mu_{g m}^{P} \mu_{m e}^{P}}{4 \hbar^{2}} \frac{\varepsilon_{P}^{*{2}}(t)}{\omega_{m g}-\omega_{P}} e^{i \Delta t} c_{e}(t)
\end{align}
In the last step, $\omega_{me}+\omega_P = \omega_m - \omega_e + \omega_P \approx \omega_m - (\omega_g + 2\omega_P) + \omega_P = \omega_{mg} - \omega_P$ relations have been used. Similarly for the coefficient for the excited state,
\begin{align}\label{excited_amp}
    i \dot{c}_{e} \approx -\sum_{m} \sum_{P} \frac{\left|\mu_{m e}^{P}\right|^{2}}{2 \hbar^{2}}\left|\varepsilon_{P}\right|^{2} \frac{\omega_{m e}}{\omega_{m e}^{2}-\omega_{P}^{2}}c_{e}(t)\notag\\
    -\sum_{m}\frac{\mu_{e m}^{P} \mu_{m g}^{P}}{4 \hbar^{2}} \frac{\varepsilon_{P}^{{2}}(t)}{\omega_{m e}-\omega_{P}} e^{-i \Delta t} c_{g}(t)\notag\\
    +\int d E c_{E}(t) e^{-i \omega_{e E} t} V_{e E}(t)
\end{align}

The amplitudes corresponding to the continuum states can be calculated using Eq.~\ref{continuum_amp}. Substituting this into the last term of Eq.~\ref{excited_amp} which corresponds to the coupling between excited state $1s2s$ and continuum state gives,
\begin{align}
    \int dE\ c_E(t) e^{-i\omega_{Ee}t} V_{eE}(t) = \frac{1}{2 i \hbar} \int dE \int_{-\infty}^{t} dt^{\prime} c_{e}\left(t^{\prime}\right) \mu_{E e}^{P}\notag\\
    \times \left[\varepsilon_{P} e^{-i\left(\omega_{P}-\omega_{E e}\right) t^{\prime}}+\varepsilon_{P}^{*} e^{i\left(\omega_{P}+\omega_{E e}\right) t^{\prime}}\right]\notag\\
    \times  \frac{\mu_{e E}^{P}}{2}\left[\varepsilon_{P} e^{-i\left(\omega_{P}+\omega_{E e}\right) t}+\varepsilon_{P}^{*}(t) e^{i\left(\omega_{P}-\omega_{E e}\right) t}\right]
\end{align}
Within RWA the terms containing detuning $\mathcal{D}_{\alpha j} = \omega_E - \omega_\alpha - \omega_j$ will be considered,
\begin{align}
    &\int dE\ c_E(t) e^{-i\omega_{Ee}t} V_{eE}(t) \notag\\
    &= \frac{1}{4i\hbar} \int d E \int_{-\infty}^t d t^{\prime} c_{e}\left(t^{\prime}\right) |\mu_{e E}^{P}|^2 |\varepsilon_P|^2e^{-i\mathcal{D}_{Pe}(t-t')}\notag\\
    &\approx \frac{1}{4i\hbar} \int d E\ \left[c_{e}\left(t\right) |\mu_{e E}^{P}|^2 |\varepsilon_P|^2 \int_{0}^\infty d T e^{-i\mathcal{D}_{Pe} T}\right]
\end{align}
where in the last step, we have used the Markovian approximation such that $c_j(t^{\prime}) \approx c_j(t)$ \cite{KNIGHT1990} and $T=t-t'$. Using the relation, 
\begin{equation}
    \int_0^{\infty} dT e^{-i\mathcal{D}_{\alpha j}T} = -i\mathcal{P}\left(\frac{1}{\mathcal{D}_{\alpha j}}\right) + \pi \delta(\mathcal{D}_{\alpha j})
\end{equation}

\begin{align}
    \int dE\ c_E(t) e^{-i\omega_{Ee}t} V_{eE}(t) = -c_{e}(t) \left[\mathcal{P} \int d E \frac{\left|\mu_{e E}^{P} |^2|\varepsilon_{P}\right|^{2}}{4 \hbar \mathcal{D}_{P e}}\right.\notag\\
    +\left.\frac{\pi \hbar\left|\varepsilon_{P}\right|^{2}}{4 i \hbar}\left|\mu_{e E=\hbar \omega_{P}+\hbar \omega_{e}}^{P}\right|^{2}\right]
\end{align}
The dynamical energy shifts $\delta E_e$ and the photoionization rate $\Gamma_e^P$ of the excited state $1s2s$ are defined in Eq.~\ref{stark_shift_continumm} and \ref{photoionization}, respectively. Substituting these relations into Eq.~\ref{excited_amp} renders Eqs.~\ref{ground_amp} and \ref{excited_amp} as a coupled set of differential equations for $c_g$ and $c_e$. These equations can be expressed in the compact matrix form given in Eq.~\ref{coupled_eqn}. The numerical solution of this equation, along with the computational details and results, is presented in Sec.~\ref{sec:numerical_and_plots}.

\section{Derivation of effective Hamiltonian for XUV-IR two-color scheme}
\label{appendix_B}

Two-photon excitation can also be realized using a two-color scheme, as described in Sec.~\ref{sec:two_color_two_photon}. However, this introduces additional pathways that couple the $1s2s$ state to the continuum through multiphoton excitation. The corresponding wavefunction is shown in Eq.~\ref{wavefunction_3level}. Substituting this into Eq.~\ref{time-dep-eq}, where $V(t)=-\bm{\mu}\cdot(\bm{E}_{1}(t) + \bm{E}_{2}(t))$, we get a new set of equations of complex amplitude similar to Eq.~\ref{eq_of_motion},

\begin{subequations}
\label{eq_of_motion2}
    \begin{align}
        i \hbar \dot{c}_{g}(t)=&\sum_{m} c_{m}(t) e^{-i \omega_{mg} t} V_{g m}(t)\\
        i \hbar \dot{c}_{e}(t)=&\sum_{m} c_{m}(t) e^{-i \omega_{m e} t} V_{e m}(t)\notag\\
        &+\int d E c_{E}(t) e^{-i \omega_{e E} t} V_{e E}(t)\\
        i \hbar \dot{c}_{b}(t)=&\sum_{m} c_{m}(t) e^{-i \omega_{m b} t} V_{b m}(t)\notag\\
        &+\int d E c_{E}(t) e^{-i \omega_{b E} t} V_{b E}(t)\\
        i \hbar \dot{c}_{m}(t)=&\sum_{j=g,e,b} c_{j}(t) e^{i \omega_{mj} t} V_{mj}(t)\\
        i \hbar \dot{c}_{E}(t)=&c_{e}(t) e^{i \omega_{E e} t} V_{E e}(t) + c_{b}(t) e^{i \omega_{E b} t} V_{E b}(t)
    \end{align}
\end{subequations}
where $\bm{E}_{1(2)}$ denotes the electric field of the XUV(IR) pulse. Rapidly oscillating nonresonant amplitudes can be obtained using adiabatic elimination, as discussed in Sec.~\ref{appendix_A}. A similar derivation is performed to obtain the amplitudes of state $|g\rangle$,
\begin{align}\label{ground_amp2}
    i \dot{c}_{g} \approx -\sum_{m} \sum_{P} \sum_{\alpha=1,2} \frac{\left|\mu_{m g}^{\alpha}\right|^{2}}{2 \hbar^{2}}\left|\varepsilon_{\alpha}\right|^{2} \frac{\omega_{m g}}{\omega_{m g}^{2}-\omega_{\alpha}^{2}}c_{g}(t)\notag\\
    -\sum_{m}\left[\frac{\mu_{g m}^{1} \mu_{m e}^{2}}{4 \hbar^{2}} \frac{\varepsilon_{1}^{*}(t)\varepsilon_{2}^{*}(t)}{\omega_{m g}-\omega_{1}} + \frac{\mu_{g m}^{2} \mu_{m e}^{1}}{4 \hbar^{2}} \frac{\varepsilon_{2}^{*}\varepsilon_{1}^{*}}{\omega_{m g}-\omega_{2}}\right]\notag\\
    \times e^{i \Delta_{eg} t} c_{e}(t)
\end{align}
where $\Delta_{eg}=(\omega_1+\omega_2)-\omega_{eg}$ is the two-photon detunings. A similar relation can also be derived for the time evolution of the amplitudes $c_e$ and $c_b$. The dynamical stark shift of the $j^{th}=\{g,e,b\}$ states corresponding to the XUV(IR) pulse is given by
\begin{align}
    S_{j}^{1(2)}(t) = -\sum_{m} \frac{\left|\mu_{m j}^{1(2)}\right|^{2}}{2\hbar^{2}}\left|\varepsilon_{1(2)}\right|^{2} \frac{\omega_{m j}}{\omega_{m j}^{2}-\omega_{1(2)}^{2}}
\end{align}
The two-color two-photon Rabi frequencies for two possible pathways are defined as
\begin{equation}
\label{rabi_freq2}
    \Omega_{ge}^{12(21)} = -\sum_{m}\frac{\mu_{e m}^{1(2)} \mu_{m g}^{2(1)}}{2 \hbar^{2}} \frac{\varepsilon_{1(2)}\varepsilon_{2(1)}}{\omega_{m g}-\omega_{1(2)}}
\end{equation}
The equation for $\dot{c}_e$ also includes contributions from the dynamical energy shifts and photoionization of the $1s2s$ state ($|e\rangle$), arising from its coupling to the continuum via the XUV pulse only.
\begin{align}
    \label{stark_shift_continumm_XUV}
    \delta E_e^1 &= - \mathcal{P}\int d E \frac{1}{4\hbar}\frac{\left|\mu_{e E}^{1}|^2 |\varepsilon_{1}\right|^{2}}{\omega_E - \omega_1 - \omega_e}\\
    \label{photoionization_XUV}
    \Gamma_{e}^1(t)&=\frac{\pi}{2 \hbar} \left|\varepsilon_{1}\right|^{2}\left|\left\langle e\left| \mu \cdot \hat{e}_{1}\right| E=\hbar \omega_{1}+\hbar \omega_{e}\right\rangle\right|^{2}
\end{align}
However, the evolution of the amplitude $c_b$, corresponding to the $1s4s$ state, includes both the dynamical energy shift and photoionization induced by the IR pulse only.
\begin{align}
    \label{stark_shift_continumm_IR}
    \delta E_b^2 &= - \mathcal{P}\int d E \frac{1}{4\hbar}\frac{\left|\mu_{b E}^{2}|^2 |\varepsilon_{2}\right|^{2}}{\omega_E - \omega_2 - \omega_b}\\
    \label{photoionization_IR}
    \Gamma_{b}^2(t)&=\frac{\pi}{2 \hbar} \left|\varepsilon_{2}\right|^{2}\left|\left\langle b\left| \mu \cdot \hat{e}_{2}\right| E=\hbar \omega_{2}+\hbar \omega_{b}\right\rangle\right|^{2}
\end{align}
Finally, the states $|e\rangle$ and $|b\rangle$ are coupled via the IR two-photon Rabi frequency,
\begin{equation}
\label{rabi_freq_IR}
    \Omega_{eb} = -\sum_{m}\frac{\mu_{b m}^{2} \mu_{me}^{2}}{2 \hbar^{2}} \frac{\varepsilon_{2}^{2}(t)}{\omega_{m e}-\omega_{2}}
\end{equation}
Combining all contributions, we arrive at Eq.~\ref{coupled_eqn_two_color}.

\bibliography{apssamp}

\end{document}